\documentclass[graybox]{svmult}

\usepackage{mathptmx}       % selects Times Roman as basic font
\usepackage{helvet}         % selects Helvetica as sans-serif font
\usepackage{courier}        % selects Courier as typewriter font
\usepackage{type1cm}        % activate if the above 3 fonts a
\usepackage{amsmath,natbib}
\usepackage{makeidx}         % allows index generation
\usepackage{graphicx}        % standard LaTeX graphics tool
\usepackage{multicol}        % used for the two-column index
\usepackage[bottom]{footmisc}% places footnotes at page bottom

\makeindex             % used for the subject index
\newcommand\ion[2]{#1$\;${\scshape{#2}}}%              
\newcommand{\e}            {\mbox{$^{-1}$}}
\newcommand{\ee}           {\mbox{$^{-2}$}}

\newcommand{\kms}          {\mbox{${\rm km~s^{-1}}$}}
\def\twelveCO{\mbox{$^{12}$CO}}
\def\13CO{\mbox{$^{13}$CO}}
\def\C18O{\mbox{C$^{18}$O}}
\def\farcs{\hbox{$.\!\!^{''}$}}

\begin{document}

\title*{The determination of protoplanetary disk masses}
% Use \titlerunning{Short Title} for an abbreviated version of
% your contribution title if the original one is too long
\author{Edwin A. Bergin and Jonathan P. Williams}
% Use \authorrunning{Short Title} for an abbreviated version of
% your contribution title if the original one is too long
\institute{Edwin A. Bergin\at Department of Astronomy, University of Michigan, 1085 S. University Ave, Ann Arbor, MI 48109 \email{ebergin@umich.edu}
\and Jonathan P. Williams\at Institute for Astronomy, University of Hawaii at Manoa, Honolulu, HI 96822 \\\email{jw@hawaii.edu}}
%
% Use the package "url.sty" to avoid
% problems with special characters
% used in your e-mail or web address
%
\maketitle

%\abstract*{}
\abstract{In this article we review the methods used to determine the gas and dust masses of protoplanetary disks, with an emphasis on the lesser characterized total gas mass.   Our review encompasses all the indirect tracers  and the methodology that is be used to isolate the hidden H$_2$ via dust, CO, and HD emission.   We discuss the overall calibration of gaseous tracers which is based on decades of study of the dense phases of the interstellar medium.    At present, disk gas masses determined via CO and HD are (in a few instances) different by orders of magnitude, hinting at either significant evolution in total disk mass or in the CO abundance .  Either of these would represent a fundamental physical or chemical process that appears to dominate the system on $\sim$million year timescales.   Efforts to reconcile these differences using existing and future facilities  are discussed.  }

\section{Introduction}
\label{sec:intro}
\begin{figure}
\sidecaption[t]
\includegraphics[width=2.8in]{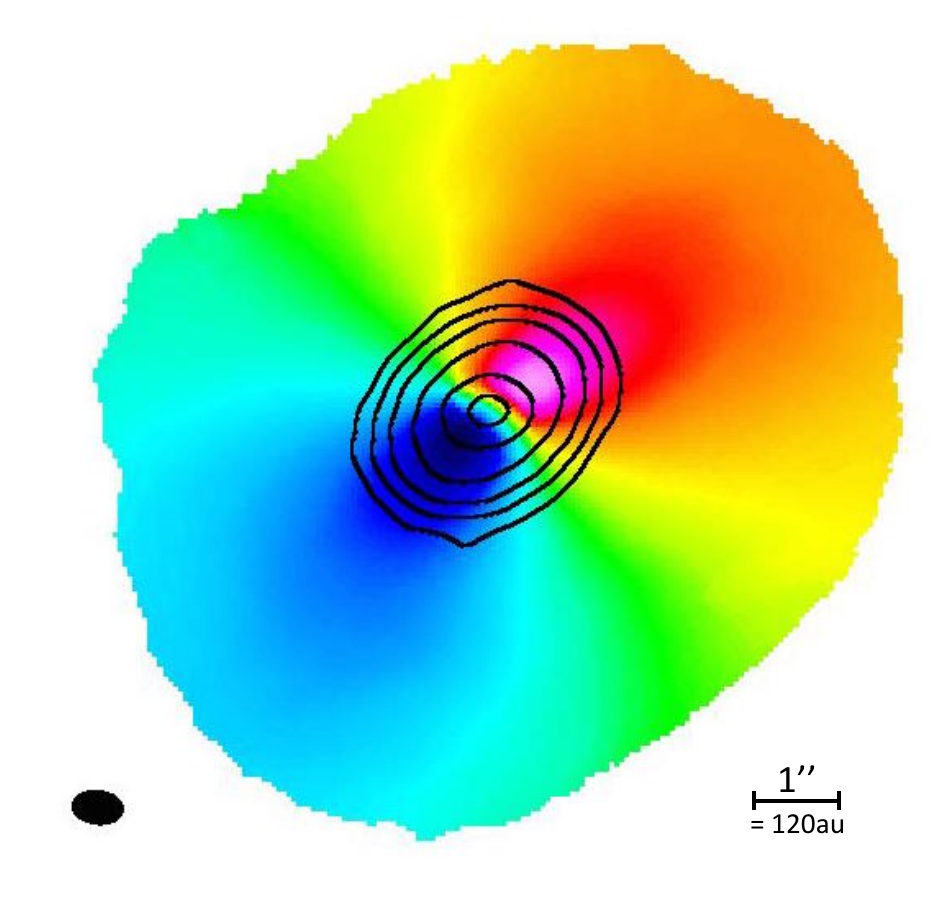}
\caption{Dust and gas in the HD 163296 protoplanetary disk.
These publicly available ALMA Science Verification data show the
$870\,\mu$m continuum emission from dust as contours and
a color-coded first moment map of CO 3--2 emission
(blue--red at 2--10\,\kms).
The elliptical $\sim 0\farcs 5\times 0\farcs 7$ beam size
and a physical scale bar is shown in each of the lower corners.
The more compact continuum emission is due to the drift of
dust grains and demonstrates the necessity to measure the
gas and dust properties independently from each other.}
\label{fig:hd163296_gas_dust}
\end{figure}

Disks are an inevitable byproduct of stellar birth and the sites of
planet formation. The determination of protoplanetary disk masses is
fundamental for understanding almost all aspects of disk physics,
from formation and evolution, to the types and compositions of
planetary outcomes. As with the molecular cores from which they
originate, disks consist of gas and dust. Unlike these cores and
the general Interstellar Medium (ISM), however, most of the solid mass
is contained in millimeter and larger sized dust grains that are not
well-mixed with the gas (Figure~\ref{fig:hd163296_gas_dust}).
Measuring disk masses therefore requires
accounting for each component separately and different techniques
are used both for the observations and modeling.    

Knowledge of the dust mass, and the evolution of dust particles, 
is central to understand the beginnings (and endings)
of planet formation.  However,  it is the gas, as the dominant
constituent, which sets the stage for understanding key aspects of the
physical and chemical evolution - and the disk {\em gas} mass is the least characterized quantity.
Today with the operation of the Atacama Large Millimeter Array (ALMA) we are
obtaining resolved images of the planet-forming zones with optically thin
tracers that probe nearly the entire column of gas perhaps, in some cases,
into the dust rich midplane where planets are born.  In the coming decade
the James Webb Space Telescope will offer an unprecedented  view of disk
systems in the mid-infrared tracing hot gas closer to the star and warmer
material at greater distances.  In all, a key goal is to use
molecular emission and dust thermal emission to set deterministic constraints on their coupled physical and chemical
evolution.  In the case of molecular hydrogen, which is the most abundant species, this relates to the timescale of gas disk
dissipation as a gas-rich system is needed for the formation of Jovian worlds; in contrast
super-Earths and ice giants could potentially form in a dissipating gas
disk.  Furthermore, knowledge
of the gas mass, and its distribution, opens the ability to use the emission maps and determine the gas
phase chemical composition, which has implications for the resulting composition of both
gas giants \citep{omb11} and terrestrial worlds \citep{Bergin15}.  
More broadly it is the interplay between the gas disk and forming planets that can
drive planetary migration and  circularize planetary orbits \citep[][and references therein]{Baruteau16}.
Thus, the ability to connect our knowledge of planetary birth  to the exploding
field of detection and characterization of exoplanets is greatly aided by grounding knowledge
of disk gas and dust masses.

It is the goal of this contribution to summarize the methodology  of how masses,
both gas and dust, are determined for protoplanetary disks.   The main focus of this
paper is on the lesser characterized gas mass, but briefly,  in \S~\ref{sec:dust}
we outline the methodology for the determination of dust masses, while in \S~\ref{sec:gas} we discuss the
various methods to approach the determining the mass of hidden H$_2$.   Thus we first
show why H$_2$ cannot be detected in the bulk of the mass and how other tracers,
such as CO or HD, can be used as calibrated tracers of the overall mass and even its
distribution within young planet-forming disk systems.   However, at present these disparate methods are finding significant
(orders of magnitude) differences in the estimated mass.   This hints that we have discovered a fundamental 
effect which is either related to fast (million year) gas disk dissipation or the formation
of icy bodies that lock volatiles, such as water and CO, into the dust rich miplane, or perhaps both.
We end with a discussion of future prospects.

\section{The determination of dust masses}
\label{sec:dust}
Even though the dust is a minor component, at least initially,
in protoplanetary disks, it is the easiest to detect because it emits over
a continuum rather than the narrow spectral lines from gaseous species.
Consequently, there have been more studies of the dust content in
disks and there are several recent reviews that describe the current
status in the field \citep{williams_araa, Testi14, Andrews15}

The intensity of the dust emission depends on its temperature and optical depth.
The latter is determined by multiplying the projected surface density,
$\Sigma_{\rm dust}$, times an opacity, $\kappa$,
that quantifies the absorption cross-section per unit mass.
Mie theory tells us that particles interact most strongly with
radiation at wavelengths comparable to their size.
Interstellar reddening shows that dust grains range from sub-micron
to micron sizes in the ISM and, at wavelengths greater than the
maximum grain size, the combined opacity can be approximated by a
power law,
\begin{equation}
\kappa_\nu = \kappa_0 \left(\frac{\nu}{\nu_0}\right)^\beta
\end{equation}
\citep{Draine06}.
The normalization and index depend on the grain composition and size
distribution \citep[e.g.,][]{pollack94, ossenkopf_henning}.
As we will see, there are many other uncertainties in determining
disk masses so here we simply use a basic parameterization from
\citet{bs90},
one of the first papers to discuss millimeter wavelength observations of disks,
$\kappa_0 = 10\,{\rm cm}^2\;{\rm g}^{-1}, \nu_0 = 1000\,{\rm GHz}, \beta=1$.

Because of the rising dependence of $\kappa_\nu$ on frequency,
disks radiate most strongly in the mid- and far-infrared.
%$\lambda\sim 10-100\,\mu$m.
The most sensitive surveys for disks, and therefore the
best statistics and constraints on their lifetime, have been
made with the Spitzer and Herschel telescopes \citep{evans09}.
However, at these high frequencies, the emission is saturated
(optically thick) and largely independent of the surface density.
At millimeter wavelengths, $\nu\sim 300$\,GHz,
the emission becomes optically thin for surface densities
$\Sigma_{\rm dust} < 1/\kappa_\nu \sim 3$\,g\,cm\ee.
This condition holds for most disks, although the inner tens of au
may exceed this in massive disks. 
The dust mass is therefore directly proportional to the
millimeter flux, $F_\nu$. For a constant temperature, $T_{\rm dust}$,
it is straightforward to derive
\begin{equation}
M_{\rm dust} = \frac{F_\nu d^2}{\kappa_\nu B_\nu(T_{\rm dust})},
\label{eq:dustmass}
\end{equation}
where $d$ is the distance to the source and $B_\nu$ is
the Planck function.
More detailed models can be made that consider the radiative
transfer and temperature structure, but the results do not greatly
differ from this simple prescription.
An additional uncertainty is the distance, $d$,
but GAIA provides much needed precision,
at least for optically bright sources.

It is also important to keep in mind that any measurement of a dust
mass is a measure of the amount of particles with sizes
comparable to the observing wavelength.
Millimeter and larger-sized particles settle toward
the midplane and drift toward the center at different rates such
that they size-segregate in a disk. Observations at different
wavelengths probe not only a different range of grain sizes but
also different regions in the disk.
This can be exploited to study disk physics, but it can also
complicate the measurement of disk dust masses
and must be taken into account when comparing results in the literature.

The process of planet formation begins with the agglomeration
of small dust grains, from sub-micron sizes in the ISM to meters and beyond.
This process appears to be remarkably fast, as evidenced from
young ages for differentiated asteroids in the Solar System
\citep{kleine02}.
At least until self-gravity becomes important,
the number distribution of particle sizes, $a$,
that evolve through agglomeration and collisions
is expected (and observed over a limited range of sizes)
to follow a power law, $N(a)\propto a^{-p}$,
with index $p\simeq 3.5$ \citep{Testi14}.
This has the interesting property that the total surface
(emitting) area, $\propto\int a^2 N(a) da$, is dominated by
the smallest grains but the total mass, $\propto\int a^3 N(a) da$,
is mostly within the largest particles.
This is the reason why disks are easier to detect than planets
and it means that the millimeter derived dust mass is a
lower limit to the total mass of solids in a disk.

Nevertheless, despite these caveats, two clear trends emerge
from millimeter wavelength surveys of dust masses.
First, for a given age, higher mass stars tend to have more massive
disks than low mass stars.
Second, disk masses decline rapidly with age such that few disks
are detectable in the millimeter beyond a few Myr.
We elaborate on each of these points below.

\subsection{Dust mass dependence on stellar mass}
To study the dependence of disk masses on stellar or other properties
requires surveys that are minimally biased.
Disks radiate most strongly in the infrared and Spitzer carried out
complete surveys of all star-forming regions within 200\,pc \citep{evans09}.
These showed that disk lifetimes are longer around lower mass stars.
Mass measurements require detecting the optically thin, and therefore
weaker, millimeter emission.
Before ALMA, disk mass surveys were painstakingly slow and only the
nearby Taurus star-forming cloud had been systematically surveyed
to high sensitivity \citep{Andrews13}.

The Taurus survey, mainly carried out with the SMA at 1.3\,mm to
an rms of about 1\,mJy, included 227 protostars (infrared Class II objects)
ranging in mass from $\sim 0.01\,M_\odot$ to $\sim 3.0\,M_\odot$.
Even though there were many non-detections, the large sample source
and high sensitivity showed the disk mass scales approximately linearly
with the host stellar mass with a median ratio,
$M_{\rm dust}/M_{\rm star}\simeq 3\times 10^{-5}$.
For sun-like stars, this corresponds to about $10\,M_\oplus$
which is about a factor of 3 lower than the total mass of elements
heavier than Hydrogen and Helium in the Solar System and suggests
that our planetary system is more massive than most.
However, there is a large intrinsic dispersion in dust masses for any
given stellar mass of about $\pm 0.7$\,dex
(i.e., $\sim 5$ times higher or lower).
Some of this may be attributed to uncertainties
in the stellar mass measurements, some to the cloud
depth of about 30\,pc \citep{Loinard:13}, and some may be disk-disk
grain opacity variations,
but it is clear that disks around stars with similar masses and ages
can have substantially different amounts of dust, probably due to
variations in their initial conditions. The general trend that higher
mass stars have more massive disks is a key assumption in population
synthesis models that predict planet mass -- stellar mass correlations
\citep{Howard12}.

The advent of ALMA promises complete millimeter wavelength surveys
of disks in many of the same regions surveyed in the infrared by Spitzer
and Herschel. The first large, systematic survey of the dust and gas
in a nearby, young star-forming region was carried out by
\citep{Ansdell16} in Lupus.
This 0.9\,mm survey had an rms of about 0.3\,mJy and included 89
protostars. The survey was 96\% complete for all protostars with
stellar masses $>0.1\,M_\odot$.
62 sources were detected in the continuum and the dust mass
was found to correlate with stellar mass, but with a steeper
dependence, $M_{\rm dust}\propto M_{\rm star}^{1.8}$.
They also found considerable dispersion about this relation,
although the range was smaller, $\pm 0.5$\,dex, than that seen in Taurus.

A similarly steep dust mass -- stellar mass scaling was found
in a recent ALMA survey of 93 disks in the $\sim 2$\,Myr Chamaeleon I
star-forming region \citep{Pascucci16}. They suggest the underlying
reason is not faster growth of grains beyond millimeter sizes
but rather shorter drift times in disks around low mass stars.

\subsection{Dust mass evolution}
Dust masses decline rapidly with time. Prior to ALMA, relatively
few disks had been detected at millimeter wavelengths
in regions older than $\sim 3$\,Myr
\citep{Carpenter05, Mathews12}.
Infrared excesses indicate that some dust was still present
around many stars but the weakness of the millimeter emission
implies that either the mass or the surface area has decreased
\citep{Williams12}.

To follow the evolution of the dust in disks therefore requires
sensitive observations. Moreover, large surveys are required to
distinguish trends from the inherent dispersion in initial conditions.
ALMA provides the ability to observe many faint sources quickly and
recent results reveal disk evolution  in far greater detail than before.
\citet{Barenfeld16} surveyed 75 sources un the 5--11\,Myr old
Upper Scorpius region and found that very little dust
remains (median $\sim 1\,M_{\rm moon}$) remains,
indicating that planetesimal formation is essentially complete
by this point.
The closest intermediate aged region, $\sim 3-5$\,Myr,
is $\sigma$\,Orionis but it is about 3 times more distant
than Lupus and Upper Scorpius so detection limits are higher.
\citet{Williams13} imaged the cluster with the SCUBA-2
submillimeter camera and found very few individual sources
but, through a stacking analysis, determined a median dust
mass $\sim 1.5\,M_{\oplus}$.
During the course of writing this paper, we received new data from an
ALMA survey of 92 disks that confirm this result and show that the
distribution of dust masses in  $\sigma$\,Orionis lie in
between those of Lupus and Upper Scorpius.
This suggests that dust masses decline steadily with time
rather than through some abrupt cutoff.
As the analysis of these datasets proceed, we will learn about
the evolutionary dependence on stellar mass.

The ubiquity of extrasolar planets tells us that the disks are
not dispersing but rather than we are witnessing the first steps
in the formation of planetesimals.
The implication is that most of the solid mass aggregates into
millimeter and larger sized particles within a few Myr.
This is consistent with cosmochemical studies of the Solar System
that show chondrules form from 0--3\,Myr after the first
metallic flakes (Calcium-Aluminum Inclusions; CAIs)
condensed from the hot nebular gas \citep{Connelly08}.

\subsection{Dust density profiles}
Up to now, we have discussed global dust masses obtained from total
continuum disk fluxes. Millimeter interferometry routinely achieves
sub-arcsecond resolution which allows disks to be resolved in nearby
star-forming regions. Consequently, we can measure the amount of
dust at different radii in a disk or, equivalently, its surface
density profile. The basic idea is to apply Equation~\ref{eq:dustmass}
for each pixel in the image with a radial variation in dust temperature
that can be matched to the infrared SED
\citep{Lay97, Andrews07}.

As the sensitivity of observations improved,
\citet{Hughes08} showed that an accretion disk profile with form,
\begin{equation}
\Sigma(R)=(2-\gamma)\,\frac{M}{2\pi R_c^2}
          \left(\frac{R}{R_c}\right)^{-\gamma}\,
          \exp\left[-\left(\frac{R}{R_c}\right)^{2-\gamma}\right],
\label{eq:SigmaR}
\end{equation}
matches the continuum visibilities better than a truncated
power law, $\Sigma(R)\propto R^{-\gamma}$ for $R\leq R_{\rm put}$
and $\Sigma(R)=0$ for $R>R_{\rm out}$.
The characteristic radii, $R_c$, range from a few tens
to a few hundred au and larger disks tend to be more massive
\citep{Andrews15}.
The typical resolution of pre-ALMA observations was about 50\,au
in nearby star-forming regions. The dust surface densities
generally rose toward the center but with a wide range of
peak values in the innermost beam of a few hundredths to a
few grams per square centimeter
(which, a posteriori, justifies our assumption that the
millimeter emission is optically thin).
\citet{Guillot99a} show that Jupiter and Saturn contain at least $30\,M_\oplus$
of elements heavier than Hydrogen and Helium.
The interpolated surface density values at their current orbital
radii, 5--10\,au, match the extrapolated dust surface densities
of the more massive disks.

Some disks are cleared of material within their inner few 
tens of au. Indications for this were first seen with IRAS
as a marked mid-infrared dip in a protostellar SED \citep{Strom89}
and thought to signal an inside-out clearing in the late stages
of planet formation.
The large, sensitive Spitzer surveys showed that between about 10-20\%
of disks in any given region are in this ``transition'' phase
\citep[][and references therein]{williams_araa}.
As millimeter interferometers expanded their baselines and improved
their resolution below an arcsecond, the purported cavities were directly
resolved and the depletion in the dust was determined to be at least two
orders of magnitude within the inner tens of au \citep{Andrews2011}.
In part because the inner holes may be created by massive
protoplanets, transition disks are the subject of considerable
current research (see Chapter by van der Marel in these proceedings).

Resolved observations at multiple wavelengths from the sub-millimeter
to centimeter reveal the surface density profiles of grains with
different sizes. Disk sizes are smaller at longer wavelengths
indicating that larger grains are more concentrated than smaller grains
\citep{Perez15, Tazzari16}.
This matches theoretical expectations that the drift speed at 100\,au
should increase with grain size up to about a centimeter
\citep{armitage10}.
It should then slow down for larger grains as they have longer
stopping times, but to detect and map their distribution would
require high resolution meter wavelength observations at
sensitivities that will only be achieved with the Square
Kilometer Array \citep{Wilner04}.

Drift speeds are proportional to the Keplerian rotation rate
and drift timescales are therefore dependent on the stellar mass.
This is testable with the combination of the expanded VLA and ALMA
and high resolution imaging disk imaging surveys at multiple wavelengths.

ALMA has just begun to transform our view of the dust
distribution in disks. Its tremendous leap not only in collecting
area but also in baseline length allows ultra-high resolution
($\sim 20-30$\,milli-arcsecond) imaging of the continuum emission.
This has led to instantly iconic images of multiple, narrow dust
rings in HL Tau \citep{HLTau_ALMA} and TW Hya \citep{Andrews16}.
The structure in these two systems is predominantly radial but
at least one disk shows spiral features that may be tidal
signatures of an embedded planet or density waves from
a massive, self-gravitating disk \citep{Perez16}.

Azimuthal asymmetries have been found in some large, massive
transition disks \citep{casassus13, vanderMarel13}.
Possible causes are planets, molecular snowlines,
pressure confinement, or vortices.
As these each rely on the gas properties to a varying extent,
to understand the features in the high resolution continuum
images requires understanding the amount and distribution of the gas.

\section{The determination of gas masses}
\label{sec:gas}
\subsection{H$_2$ as a Direct Probe}
\label{sec:h2}

In the interstellar medium the gas to dust mass ratio is measured to be $\sim 100$
\citep{Goldsmith97}.  Thus, at birth, the majority of the disk mass  is
found in the gas.   Measuring the amount of gas is complicated, however, because the emission from the
dominant constituent, molecular hydrogen
(H$_2$), suffers from several effects related to its particular molecular physics  \citep{Field66} that lead to its emission being
weak and/or undetectable in most regions of a disk \citep{carmona08}.   H$_2$ is a light molecule
with large energy spacings between its rotational levels in the ground
vibrational state.   Furthermore, as a homo-nuclear molecule it has no
dipole moment and only weaker quadrupole transitions with $\Delta J = \pm 2$
are allowed, where $J$ is the quantized total rotational angular momentum.
The total nuclear spin of H$_2$ is either $S = 0$ or $S = 1$ which gives rise to a singlet state ($2S + 1 = 1$, antisymetric) or triplet ($2S + 1 = 3$, symmetric) state.   The total wave function including electronic, vibrational, rotational, and nuclear spin must be antisymmetric with respect to the exchange of any two particles (Fermi-Dirac statistics).   Since the ground electronic ($X^1\Sigma_g^+$) and vibrational states are symmetric, this means the combined rotational/nuclear spin wavefunction must be antisymmetric.   Thus there are two distinct forms of H$_2$: para-H$_2$ ($S =1$; singlet) where J must be even/symmetric (0, 2, 4, etc.) and ortho-H$_2$ ($S=3$; triplet) with J odd/anti-symmetric (1, 3, 5, etc.).   These two species cannot be exchanged by normal inelastic collisions, nor are they connected via radiative transitions \citep{dennison27}.  

Thus the fundamental ground state transition of molecular hydrogen is the
$J = 2 \rightarrow 0$ or S(0) with an
energy spacing of 510 K at 28.2 $\mu$m.    It is this energy spacing combined with the weaker quadrupole,  that makes it difficult for H$_2$ to emit appreciably in cold (T $\sim$ 20~K) molecule-dominated regions such as protoplanetary disks.   This requires H$_2$ emission to be found, if at all, in the inner regions of the disk closer to the star where temperatures are much higher ($> 100$~K).
However, the large dust column densities in the inner tens of AU of a typical protoplanetary disk imply high dust optical  depths at 28 $\mu$m \citep{pascucci_feps}.  Thus, the detection of this line is more difficult because the warm gas closer to the star where the ground state line of  H$_2$ can be excited co-exists with the optically thick dust layers.

This is essentially why stable and abundant molecules, particularly CO, have been used as gas-phase proxies to determine the physical properties of H$_2$ (density, temperature, mass, velocity field) in regions of star and planet formation.  This can be demonstrated quite readily.  For optically thin emission the line intensity is:

\begin{equation}
I_\nu = \frac{f_u N_{tot} A_{ul}hv}{4\pi}.
\end{equation}
\noindent Where $f_u$ is the fractional population.  In LTE this can be approximated as $f_u = g_u exp(-\Delta E/kT)/Q(T)$;  $g_u$ is the degeneracy and $Q(T)$ is the partition function.   The ratio of the line intensity of the ground rotational state of CO to that of H$_2$ can be expressed as:

\begin{equation}
\frac{I_{\rm CO\;1-0}}{I_{\rm H_2\; 2-0}} = \frac{f_{\rm CO, J=1}(T) N_{CO} A_{CO\;1-0}v_{CO\;1-0}}{f_{\rm H_2, J=2}(T) N_{H_2} A_{H_2\;2-0}v_{H_2\;2-0}}.
\end{equation}
\noindent For reference $A_{\rm H_2 \;2-0} = 2.94 \times 10^{-11}$~s$^{-1}$, $\nu_{\rm H_2 \;2-0} = 10.6$~THz, $A_{CO\;1-0} = 7.24 \times 10^{-8}$~s$^{-1}$, $\nu_{CO\;1-0} = 115.27$~GHz, and $N_{CO}/N_{H_2}$ is generally 10$^{-4}$ (\S~\ref{sec:gas_cal}).   Putting in these constants and assuming 20 K ($f_{\rm CO, J=1} = 0.55; f_{\rm H_2, J=2} = 4.2 \times 10^{-11}$) we find:

\begin{equation}
\frac{I_{\rm CO\;1-0}}{I_{\rm H_2\; 2-0}} = 3.5 \times 10^{7}.
\end{equation}
This quite readily brings the point home.  For cold gas at 20 K, which is typical of much of the mass, CO rotational lines will be ten million times more emissive.  Combined with issues regarding dust optical depth in the infrared, we must use other {\em calibrated} tracers.

 \begin{figure}
\includegraphics[width=4.5in]{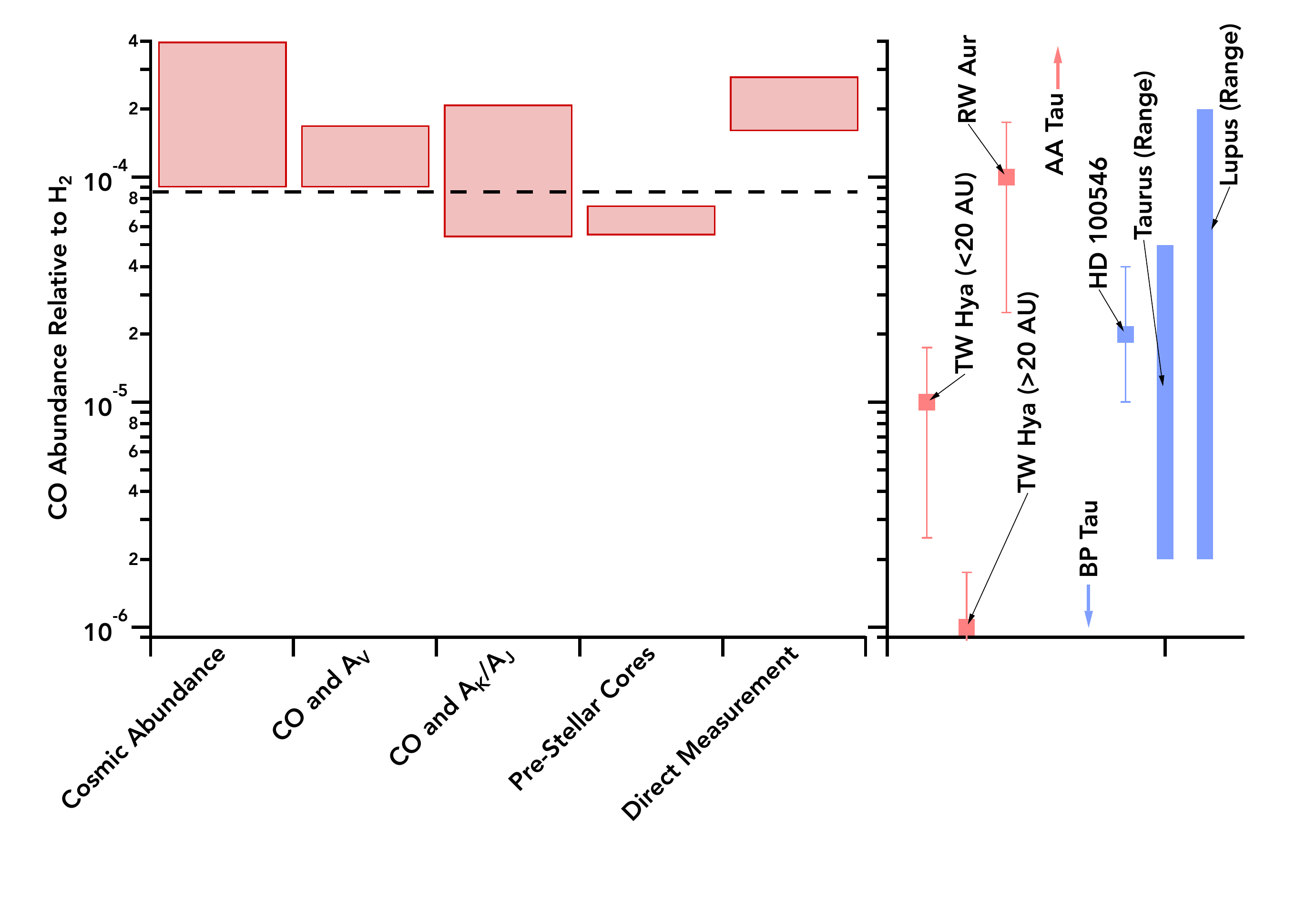}
\caption{{\em Left:} Abundance estimates of carbon monoxide in the interstellar medium with references given in \S~\ref{sec:gas_cal}.   The dashed line shows the commonly assumed reference value from Taurus \citep{flw82}.  {\em Right:} CO abundance measurements in protoplanetary disks with references similarly given in \S~\ref{sec:gas_cal}.  Values given in red derive the H$_2$ column from gas phase H$_2$or HD, while the blue points determine the H$_2$ mass from the dust assuming a dust mass opacity coefficient and a gas/dust mass ratio of 100.}
\label{fig:abuncal}
\end{figure}

\subsection{Calibration of Tracers}
\label{sec:gas_cal}

\subsubsection{Carbon Monoxide}

A critical point in using optically thin lines of various molecules to probe H$_2$ is the abundance must be known to reasonably high precision.  Furthermore the molecule must be chemically stable such that there are strong reasons to expect its abundance will be robust under a wide range of conditions.   Both of these factors lead towards carbon monoxide,  It is widely abundant in the interstellar medium (ISM) and it is expected to be provided to the disk at or near its ISM abundance.     In this section we explore this calibration.

Thus in Fig.~\ref{fig:abuncal} we provide various estimates of the $^{12}$CO abundance in the ISM (on the left-hand side) which is the motivation for its use as a calibrated mass tracer.   First we should state that the emission of $^{12}$CO is generally optically thick in the ISM; instead observations of isotopologues at lower abundance are used (e.g. $^{13}$CO or C$^{18}$O).  Here we have placed all values on the same scale by multiplying by an isotopic ratio.

We start with the cosmic abundance of carbon.  Very early on chemical models of dense gas demonstrated that the a primary product of ion-molecule chemistry would be carbon monoxide \citep{hk73}.  Thus any available carbon atoms/ions would, in short order ($<10^5$~yrs at $n_{\rm H_2} = 10^4$~cm$^{-3}$), find their way into CO, particuarly given the fact that oxygen has greater cosmic abundance when compared to carbon \citep{asplund09}.   In the diffuse ISM, CO is (mostly) dissociated, and all cosmic carbon resides in ionized form due to the 11.2 eV ionization potential of the carbon atom.   Furthermore, in this gas one can measure hydrogen directly via Ly $\alpha$ for H I or the Lyman-Werner bands of H$_2$.  Thus the abundance of carbon available to form CO can be measured.  \citet{Parvathi12} provide a comprehensive look into the carbon abundance over numerous sight lines, and this provides the range in the ``cosmic abundance'' shown in Fig.~\ref{fig:abuncal}   We note that a fraction of carbon is found to be present in the solid state \citep{Mishra15}.  Thus one cannot use the solar abundance for this calculation without subtracting the abundance of carbon locked in dust.

The next sets of measurements come from molecular clouds.  Here the dust is optically thick in the ultraviolet and one must approach the question of the H$_2$ column density via other means to determine the H$_2$ abundance.  This is traditionally done via the separate calibration of \ion{H}{I} and H$_2$ vs extinction (A$_V$).  This was first performed with {\em Copernicus} by \citet{Bohlin78}, and reconfirmed by {\em} FUSE \citep{Rachford09}, to be  N(\ion{H}{I} + 2H$_2$)$ \simeq  1.9 \times 10^{21}$ cm$^{-2}$ $A_V$.\footnote{This assumes a standard galactic extinction curve as the calibration is strictly using E(B-V).}   Thus \citet{Dickman78} observed several lines of sight with bright background stars in the vicinity of molecular clouds and explored the correlation of $^{13}$CO against  extinction (and by proxy H$_2$; meaning that the slope of the line is the molecular abundance).   This was extended to studies using C$^{18}$O \citep{flw82}, giving the range shown as CO and A$_V$.  The value from the latter study is often used as a canonical CO abundance (relative to H$_2$) of 8.5 $\times 10^{-5}$, which is shown as the dashed line in Fig.~\ref{fig:abuncal}.

More recently, with the advent of wide field near-infrared imaging, numerous studies have been able to make high fidelity extinction maps of a range of nearby clouds \citep{Kramer99, Lada94, hlh04, Pineda10, ripple13} that have been used to determine the CO abundance (CO and A$_J$/A$_K$ in Fig.~\ref{fig:abuncal}).   Here there is slightly greater fidelity in comparison to the molecular observations.   The earlier measurements discussed in the previous paragraph compared (for the most part) pencil beam observations towards a background star to much larger angular scale observations of molecular emission.   In the larger scale near-infrared imaging surveys, there are observations of numerous background sources, each with a measured extinction, within a molecular beam of 10-30$''$. These can be convolved with a Gaussian beam of comparable dimensions to the molecular observations giving a direct comparison of A$_J$ or A$_K$ to the molecular emission.  The calibration requires an additional step as the measurement determines A$_J$ or A$_K$, which is scaled to A$_V$ based on the assumed extinction curve \citep[e.g.,][]{fitzpatrick04} and then to the hydrogen column.  Attempts have also been made to calibrate this more directly \citep{Kramer98, bianchi_td, Suutarinen13}.   These studies cover both quiescent and star-forming gas with an abundance range from $\sim 5 \times 10^{-5}$ to $2 \times 10^{-4}$. The highest CO abundance is associated with regions of active star formation \citep{ripple13}.  This must hint at some potential loss of CO in regions in colder regions unassociated with stellar birth, with return in the warmer gas near young stars.   The likely culprit in the freeze-out of CO onto ice-coated dust grains \citep[e.g.][]{bt_araa}.

There is certain evidence for CO freezeout in measurements of of the CO abundance in centrally concentrated pre-stellar cores, such as Barnard 68 or Lynds 1544 \citep{Caselli99, bergin_b68}.   These cores have gas density profiles estimated either via extinction or dust thermal continuum emission and have been subject to detailed chemical study.   This, along with the detection of CO ice \citep{Boogert15} provide detailed evidence of the pervasive presence of molecular freeze-out towards dense core centers \citep[][and references therein]{bt_araa}.   In particular, with the physical conditions constrained as a function of position, models are able to extract chemical abundance profiles as a function of depth \citep[e.g.][]{bergin_b68, hotzel_b68}.   In general, the CO abundance profiles rise from low values at the edge, peak at visual extinctions of 1-2$^m$, and decline inwards as the increasing density leads to faster gas-grain collisions and subsequent freeze-out.   The values shown in Fig.~\ref{fig:abuncal} are taken at the peak  CO abundance from models where this information can be extracted.   The fact that these are at the low end of the distribution suggests that some CO resides on grains even near the core edges, reducing the overall CO abundance in cold ($<$ 20 K) gas.

Finally, in two instances the CO abundance has been measured directly using absorption of CO and H$_2$ from their respective ground vibrational state \citep{Lacy94, Goto15}.  These measurements are have fairly low precision (1 to 2$\sigma$).  However, they do represent the most direct measurement.

In general, it is fair to state that the overall range of CO abundances, excluding the pre-stellar case where the effects of freeze-out are prevalent, is quite narrow between CO/H$_2$ $\sim 0.5 - 3 \times 10^{-4}$.   This is quite reassuring for the calibration.  A canonical $^{12}$CO abundance of $1 - 2 \times 10^{-4}$ is entirely justified by the available information.  Of course, observations of lesser abundant isotopologues are required to lower the emission line optical depth.  Thus some additional assumptions regarding the atomic isotopic ratio are needed to transfer this calibration to molecular line observations. 

\subsubsection{Hydrogen Deuteride}

The abundance of HD can be characterized in similar fashion as above and has additional importance due to the use of the atomic deuterium abundance to constrain cosmological models \citep{Sarkar96}.  The abundance of primordial deuterium was set in the Big Bang and has decreased in time only due to processing through stars or astration \citep{Epstein76}.  
In atomic form there are ultraviolet absorption line studies of both atomic D and H towards bright stars.  These have most recently been performed by FUSE satellite with the result that the {\em atomic} D/H ratio is $1.56  \pm 0.04 \times 10^{-5}$  towards sources with log(N$_{\rm H_2}$)$ <$ 19.2 or distance $\leq$ 100 pc \citep[inside the local bubble;][]{Wood04, Oliveira06}.   Beyond this column there is greater dispersion with some lines of sight having lower values ($\sim 0.7 \times 10^{-5}$) and others with 40\% higher ratios above the local value.  On the low side this difference could be due to stellar astration or incorporation of deuterium in solids.

There are some measurements of HD and H$_2$ in both the UV (electronic states) and infrared (rotational states).   Thus \citet{Snow08} in a FUSE survey demonstrates the clear influence of photodissociation on the abundance of HD.  Hydrogen deuteride, like H$_2$, can self-shield itself from the destructive effects of UV photons, but at reduced efficiency \citep{Wolcott-green11}.  Thus, the HD/H$_2$ ratio approaches 2$\times$ the atomic ratio in denser shielded gas.
In addition, in dense photodissociation regions (PDR) or in shocks both HD and H$_2$ rotational lines have been detected, allowing for a direct abundance determination.  Thus in the Orion  Bar PDR \citep{Bertoldi99}  and in shocks \citep{Yuan12} there are abundance estimates of HD/H$_2$ $\sim 10^{-5}$.  Both of these are consistent with the atomic lines of sight, beyond the local bubble, where the D/H ratio is lower.    However, \citet{Yuan12} note that uncertainties in the excitation conditions encompass the primordial values.

In sum, the standard assumption for the HD/H$_2$ abundance is to assume the local value, but account for the fact that 2 hydrogen atoms are present in H$_2$; thus HD/H$_2$ $\sim 3 \times 10^{-5}$.  This is not only for the so far limited use of HD as a mass tracer in disks, but also for models of deuterium chemistry \citep[e.g.][]{Ceccarelli14}.  However, with the exception of TW Hya, most disk systems are found at distances outside the local bubble.   Hence there is clear uncertainty in this calibration with the possibility that the abundance is actually lower, meaning the mass measurement would be increased.   It is worth noting that the D/H ratio of HD in the atmosphere of Jupiter is near the estimated proto-solar value \citep{Lellouch99} suggesting in our solar system at least that the D was in HD (for the most part) and not sequestered within other molecular forms.

\subsection{Gas Mass Measurements using HD}\label{sec:gas_HD}

\begin{figure}
 \centering
 
\includegraphics[width=4.5in]{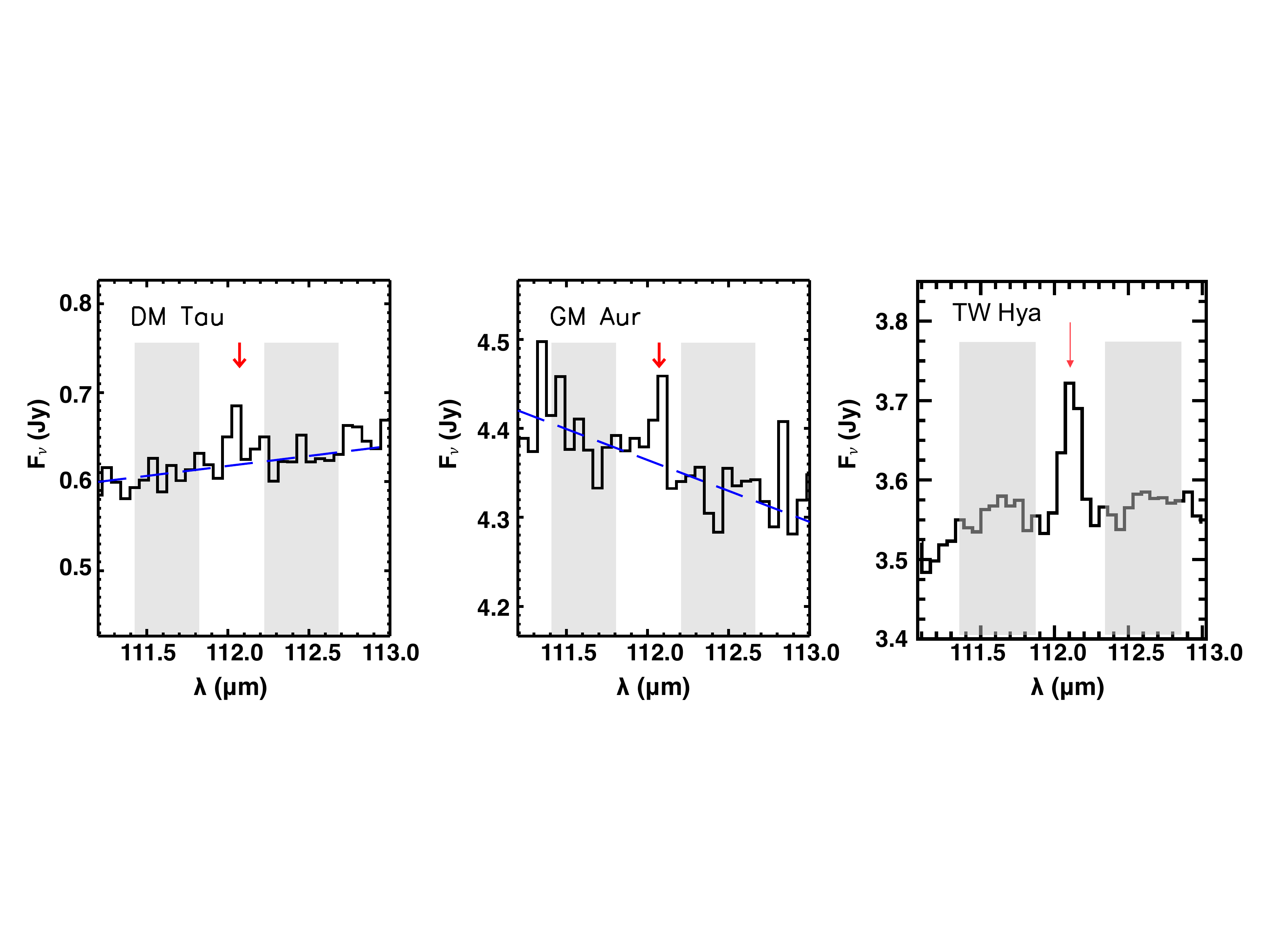}
\caption{Detection of HD J $= 1 \rightarrow 0$ towards TW Hya by \citet{bergin_hd} in the right-hand panel, along with GM Aur and DM Tau by \citet{McClure16} shown in the middle and left-hand panels, respectively.}
\label{fig:hd}
\end{figure}

Observations of HD offer an additional indirect approach to the H$_2$ mass in disk systems.   The general methodology is outlined by \citet{bergin_hd} and \citet{McClure16}.  The TW Hya detection is shown in Fig.~\ref{fig:hd}.
Since HD is an isotopologue of H$_2$ it has the same chemistry in the sense that like H$_2$, HD will not freeze onto grains as ice.  Its abundance relative to H$_2$ is well calibrated and is discussed in \ref{sec:gas_cal}.    Below we outline a general approach that illustrates how HD emission can be used as a probe of the gas mass, but also a more systematic look at dependencies.

We can derive the mass implied by an observed flux of an optically thin unresolved source  as follows.
The total number of HD molecules
($\mathcal{N_{\rm HD}}$) can be related to the line flux, 

 \begin{equation}
 F_l = \frac{\mathcal{N_{\rm HD}}  A_{10} h \nu f_{u}}{4 \pi D^2}.
 \end{equation}

  \noindent In this equation, $D$ is the distance, $\nu$  the line center
frequency, and $f_u$ is the fractional population in the upper state:
  
  \begin{equation}
f_u = 3.0 * exp(-128.5\;{\rm K}/T)/Q(T).
 \end{equation}

\noindent The total gas mass of H$_2$ is then: $M_{gas\;disk} = 2.37 * m_{\rm H} \mathcal{N_{\rm HD}}/
x_{\rm HD}$.  In this expression $x_{\rm HD}$ is the abundance of HD relative to H$_2$,
$m_{\rm H}$ is the mass of a hydrogen atom, and 2.37 is the mean 
molecular weight per particle, including helium and heavy elements \citep{kauffmann08}.
Combining these two expressions gives a relation between observed line flux and mass,

  \begin{equation}
  M_{gas\;disk} = \frac{2.37 m_H 4 \pi D^2 F_l}{A_{10} h \nu x_{\rm
HD} f_u}.
  \end{equation}

If we then include physical constants  and note that the  partition function, $Q(T)$, is near unity below $\sim 50$
K \citep{muller05}, we obtain this simple relation that shows key dependencies where $F_{TWH} = 6.3 \times
10^{-18}\;{\rm W\;m^{-2}}$ is the observed flux density toward TW Hya,

\begin{equation}
M_{gas\;disk} > 5.21 \times 10^{-5} \left(\frac{F_l}{F_{TWH}}\right)\left(\frac{3 \times 10^{-5}}{x_{\rm
HD}}\right)\left(\frac{D}{55\;{\rm pc}}\right)^2\; \exp\left(\frac{128.5\;{\rm
K}}{T_{gas}}\right) \;{\rm M_{\odot}}.
\end{equation}

\noindent The two most important factors in this expression are the abundance of HD and the sharp dependence on the gas temperature.    The abundance of atomic D is discussed in \S~\ref{sec:gas_cal}.    A bigger factor is the strong temperature dependence which requires {\em a priori} information on the gas temperature structure.
It is important to note that this estimate is a lower limit as HD emission will not trace significant mass with temperatures below 20 K. 
Another factor is the dust opacity at 112 $\mu$m; an optically thick dust-rich midplane would be hidden from HD emission. Because of these two issues this equation is listed as a lower limit.     As such the estimation of the total mass requires a model that accounts for the mass that might be hidden.    Below we illustrate how models can be used to account for the hidden material and incorporate a more realistic approximation for the gas temperature structure.

It is well known that disks are flared and directly exposed to stellar heating which leads to sharp thermal gradients with disk {\em dust} temperature having significant radial (warmer closer to the star) and vertical (warmer on the exposed disk surface) structure.    However this is only part of the picture.    Sophisticated models that include the {\em gas} thermal physics and chemistry \citep{woitke09, thi10, gorti11} have shown that midplane densities are high enough to thermally couple the gas to the dust grain, but on the disk surface the gas is more emissive in the HD line because it is directly heated by stellar irradiation and thus warmer than the dust.    In the HD detection paper, \citet{bergin_hd} adopted the thermo-chemical model of TW Hya by \citet{gorti11} to show that the HD flux was best represented by models with gas mass $>$ 0.05 M$_\odot$. Lower masses such as in inferred by \citet{thi10} of $\sim 10^{-3}$ M$_\odot$ using other tracers (e.g. $^{13}$CO, \ion{O}{i}) are inconsistent with the observed HD flux.    More recently, \citet{McClure16} published two additional HD J = 1-0 detections at the 3$\sigma$ level in DM Tau and GM Aur.   This work applied a different approach using traditional models of the dust spectral energy distribution in flared disk with warm surfaces \citep{DAlessio98}  in concert with the constraints from the HD emission to set more stringent limits on both the gas and dust mass in these systems.   

\begin{figure}
\includegraphics[width=4.5in]{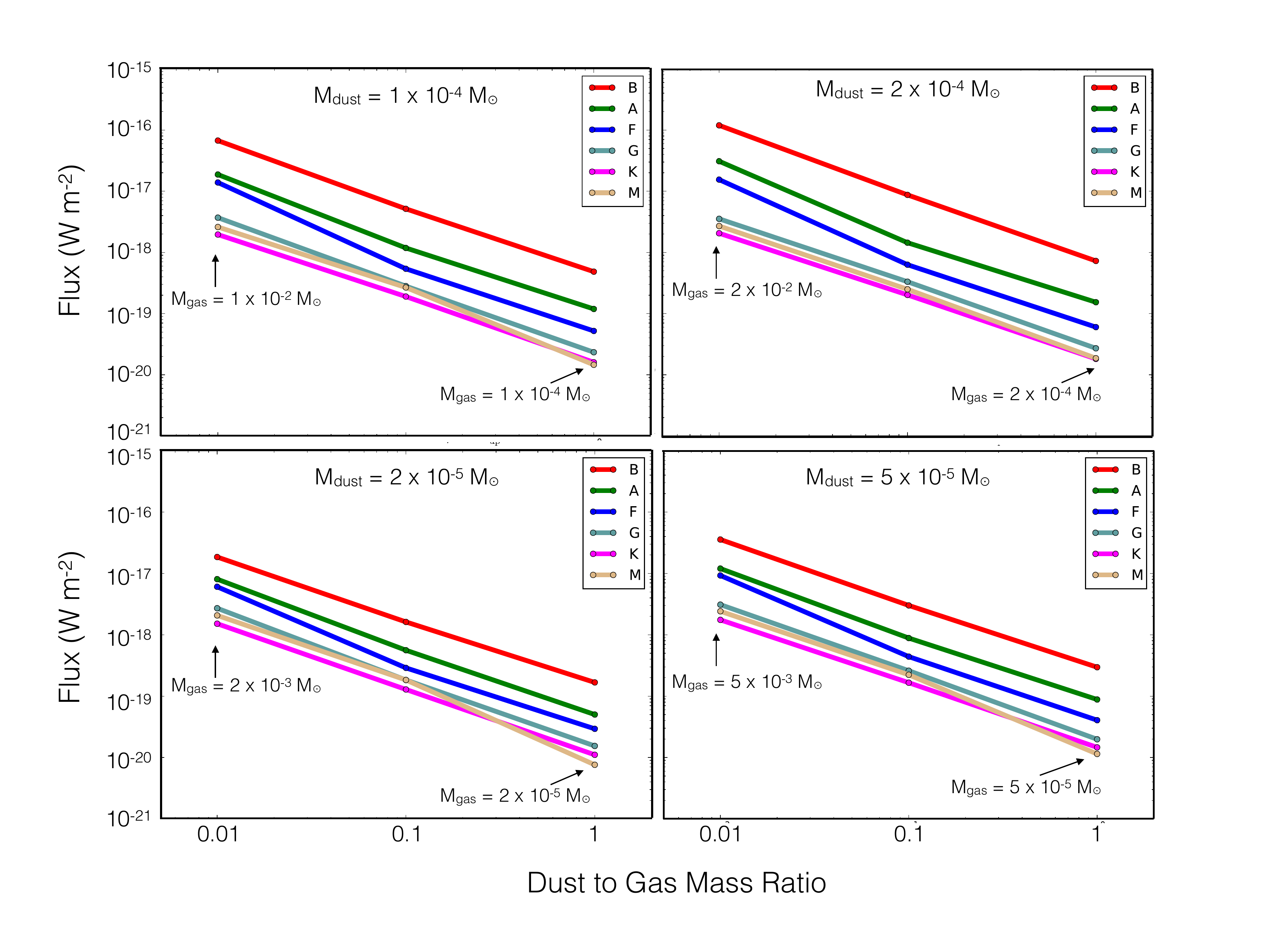}
\caption{Plot of predicted HD J = 1-0  flux density for a range of thermochemical models of \citet{Du14} and \citet{Du16} assuming different dust masses, dust-to-gas mass ratios (i.e. gas mass), and stellar spectral type.  The model disks are assumed to have an outer radius of 400 AU and are uniformly placed at 100 pc to allow for ease in distance scaling.   Most of the HD emission arises from the inner 100 AU.
}
\label{fig:hd_mass}
\end{figure}

To explore general dependencies and illustrate the utility of HD emission as a mass tracer, we employ the thermochemical model of \citet{Du14}.
The model assumes a stellar spectral type, and dust surface density distribution, along with a high energy UV radiation field motivated by observations.  For the UV radiation field, B stars are dominated by the stellar field while M stars by accretion luminosity.  Within this framework, the vertical structure is determined assuming hydrostatic equilibrium where we calculated the dust temperature and then iterated on the heating-cooling balance with the extensive chemistry to determine the gas temperature.  The models presented here were used to explore the ground state water emission \citep{Du16} and are run over a range of total disk dust mass (0.25 to 2.0 × 10$^{-4}$ M$_\odot$) that is grounded to the low end of the observed distribution of dust mass in Taurus \citep{williams_araa}.   The results, are shown in Fig.~\ref{fig:hd_mass}, and provide the predicted HD flux at 112 $\mu$m as a function of disk dust mass (labeled in each figure), the stellar type, and the dust-to-gas mass ratio.  For a fixed dust gas of the disk, a lower dust-to-gas mass ratio means a higher gas mass. 

Two effects are readily apparent in Fig.~\ref{fig:hd_mass}.  First, the effect of direct gas heating is evident as the earlier type stars (A-F) have stronger flux with greater spacing between lines for a given spectral type.   For cooler stars (G-M), at a given gas and dust mass, the predicted HD emission flux density is similar, as this is the region
the gas and dust temperatures become coupled.  Regardless of these effects, the mass dependence is clear: for a given disk dust mass and stellar spectral type, over a 2 order of magnitude range of {\em gas} mass there is a two order of magnitude increase in the HD line flux.    Clearly there are complications in this analysis as the optical depth of the line (when detected) is not constrained and there needs to be some additional information regarding the gas temperature.     However, in the latter case, resolved observations with ALMA can readily provide both radial and vertical temperature information, thereby improving our ability to use the  fundamental rotational transition of HD as a mass and gas surface density probe \citep[as demonstrated by][]{Schwarz16}.  

Finally, an additional issue regards the origin of the HD line flux.  All current data are spectrally unresolved.  If HD is emitting from a hot wind or jet component then less mass is needed to match the overall emission.   Existing models account for the entirety of the line flux as arising from the disk and, in this instance, lower masses would result.    Models certainly demonstrate that the disk will be emissive in this line, but additional spectrally resolved observations are required to confirm if current assumptions are correct.  

The high gas masses inferred from HD are roughly consistent with a disk gas-to-dust
ratio of 100, similar to their initial conditions in the ISM.
Independent evidence in support of this comes from measurements of gas accretion onto stars.
Mass accretion rates typically range from $10^{-10}-10^{-8}\,M_\odot~{\rm yr}^{-1}$,
with strong dependencies on stellar mass and age.
The total accreted mass in Lupus (and other young regions with ages of $1-3$\,Myr)
is $\sim 10^{-4}-10^{-2}\,M_\odot$ for stellar masses $\sim 0.1-1\,M_\odot$,
which is roughly consistent with the measured dust mass times 100 \citep{Manara16}.

\subsection{Gas Mass Measurements using CO}\label{sec:gas_CO}
\subsubsection{Total Mass}

An alternative approach to measuring gas masses follows the history of molecular
cloud studies and uses millimeter wavelength observations of CO and its isotopologues.
Interferometry, beginning with pioneering millimeter arrays at Hat Creek and
Owens Valley observatories in California and continuing through the present day
with ALMA, allows sensitive sub-arcsecond observations and is ideally suited
to studying the gas content and distribution in protoplanetary disks.
The first measurement of disk rotation was made through CO observations
by \citet{Sargent87}.
The emission from the primary species, \twelveCO, is so abundant
that it is optically thick and therefore its emission is more sensitive
to temperature than column density.
As with molecular clouds and cores, the lines of the rarer isotopologues
\13CO\ and \C18O, are less saturated and provide a simple pathway to
quantifying the gas mass.
However, prior to ALMA, there were surprisingly few observations of \13CO\
and almost none of \C18O. The reason can be found in the line survey of
TW Hya by \citet{Kastner97} who detect strong emission from CO, HCO$^+$, HCN, and CN,
but very weak \13CO. The bright lines are characteristic of a photon-dominated
region (PDR) but the weak \13CO\ suggests a low total gas content
(assuming the calibrated abundance discussed in \S~\ref{sec:gas_cal}).
Subsequent observations understandably tended to focus on the brighter,
more easily detectable lines.

\begin{figure}
\includegraphics[width=4.5in]{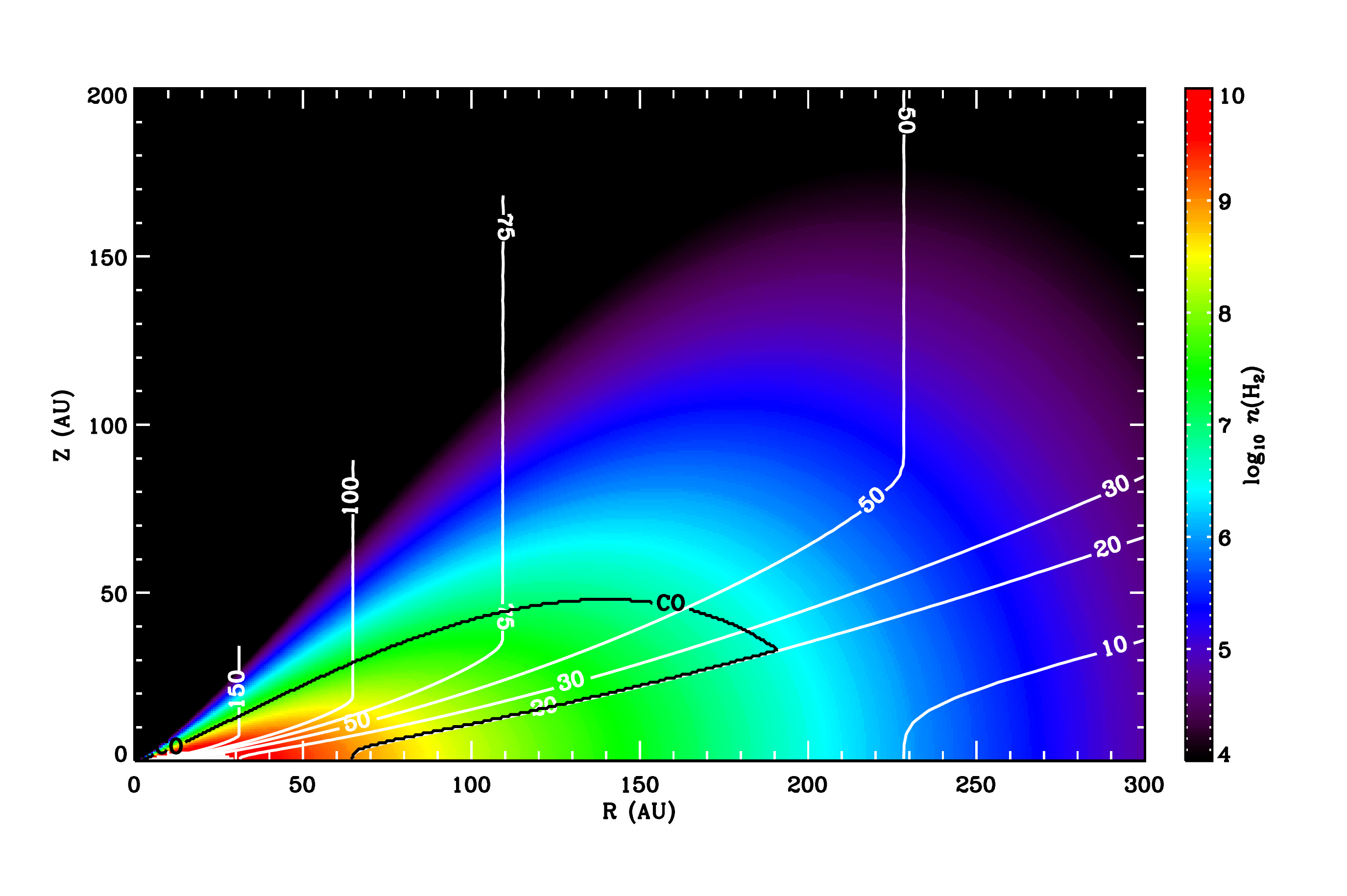}
\caption{The density and temperature distribution of a model disk.
The star is at the origin and the disk is
radially symmetric with mirror symmetry about the midplane at $Z = 0$.
The color scale represents the H$_2$ gas density on a logarithmic scale.
The gas temperature is shown and labeled in the white contours.
The black contours, labeled ‘CO’, represent the boundary of a warm
molecular layer within which CO is expected to be in the gas phase
and emit millimeter wavelength rotational lines.}
\label{fig:disk_schematic2}
\end{figure}

As our knowledge of protoplanetary disk structure and chemistry grew,
\citet{aikawa_vanz02} postulated that CO and other molecules should be found in a
{\it warm molecular layer} with a lower boundary set by the freeze-out
of molecules onto dust grains in the cold disk midplane and an upper
boundary set by photo-dissociation from the central star or external
radiation field (Figure~\ref{fig:disk_schematic2}).
Disk rotation allows us to tomographically map disk structure
and such a chevron pattern of emission has been directly observed in
the CO 3--2 line toward the HD\,163296 disk \citep{Rosenfeld13}.
For a very cold or low-density disk, the CO emitting region could be
so small as to explain the weak \13CO\ emission.
However, over the range of parameters expected for the temperature structure
and disk size based on theory and observations,
\citet{Williams14a} showed that most of the gas mass does indeed lie in the warm
molecular layer and therefore that the CO isotopologue lines should
be a reliable measure of the total disk gas content.

\begin{figure}
\sidecaption[t]
\includegraphics[width=2.8in]{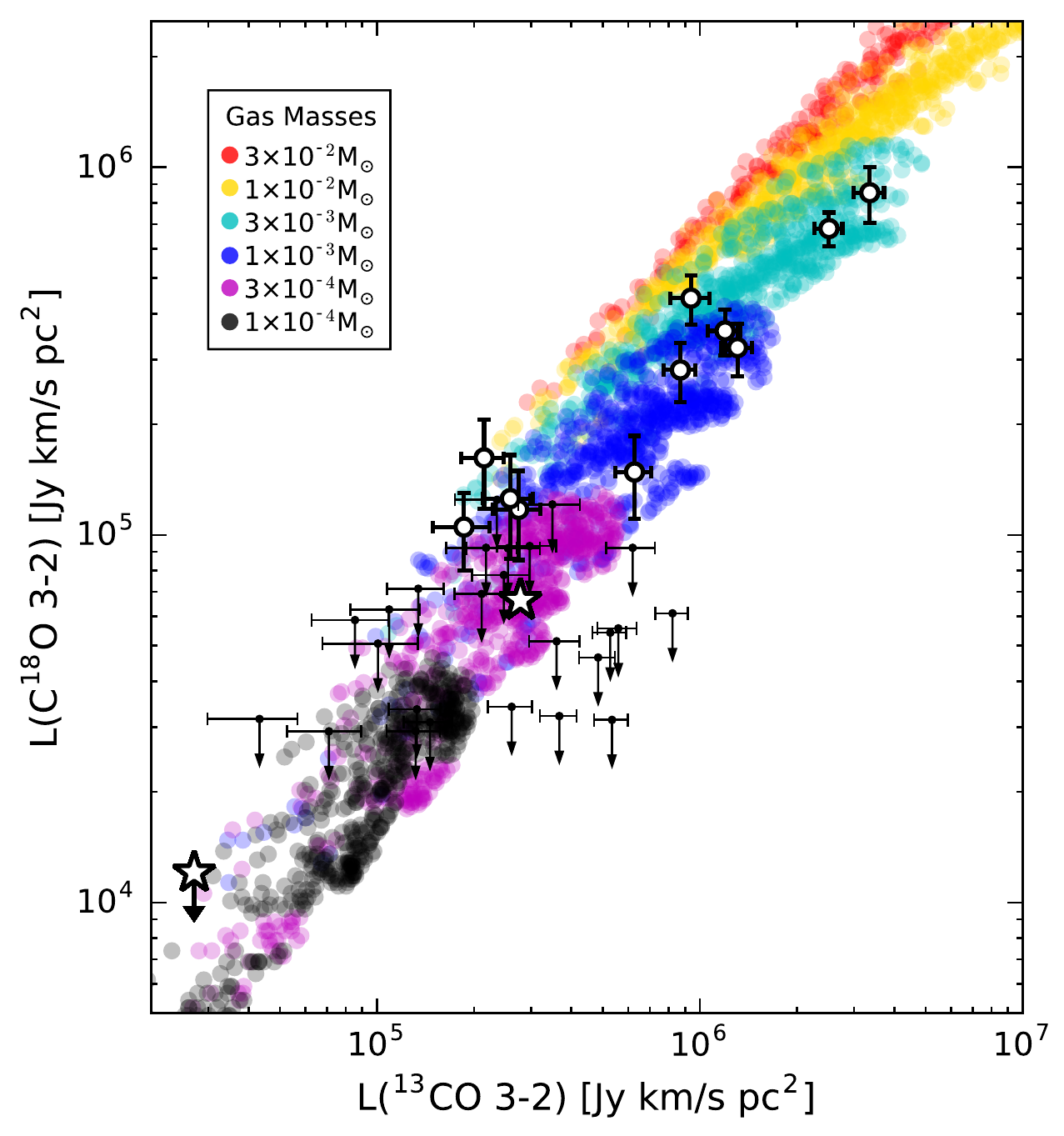}
\caption{
\C18O\ 3--2 vs. \13CO\ 3--2 line luminosity for models from the
WB14 grid, color-coded by disk gas mass. The spread for each mass
is due to optical depth, excitation, and variation in the gas
fraction within the CO-emitting warm molecular layer over the
range of disk density and temperature profiles in the models.
Nevertheless, there is sufficient separation in this isotopologue
plot that the total disk gas mass can generally be determined to
within a factor of 3--10. Observations from the \citet{Ansdell16}
Lupus survey are overplotted.
}
\label{fig:lupus_CO_luminosities}
\end{figure}

With the basic assumptions of Keplerian rotation, hydrostatic equilibrium, a constant gas-phase CO abundance above the freeze-out dominated midplane,
and azimuthal asymmetry, the spectral line profiles of CO and its isotopologues
can be calculated with radiative transfer models for any given temperature structure,
\citet{Williams14a} ran a large grid of models over a range of temperatures and densities
and found a simple way to measure disk masses by plotting the
\13CO\ and \C18O\ line luminosities against each other as in
Figure~\ref{fig:lupus_CO_luminosities}.

The luminosities of the two isotopologues closely correlate of course,
but disks of a fixed mass can have a wide range of values depending on
molecular excitation, line optical depth, and the amount of CO freeze-out
and photodissociation. The constraints from two lines helps to remove the
ambiguities and, despite the  intrinsic scatter, the spatially and velocity
integrated line luminosities provide a simple and reliable measure of the
total gas mass, to within a factor of 3--10.
This simple procedure is ideally suited for the statistical analysis
of large datasets.
Figure~\ref{fig:lupus_CO_luminosities} shows observed values from
the Ansdell ALMA Lupus survey and the good match with this simple
model description. For each data point, the mass or upper limit can be
simply read off this plot.

\begin{figure}
\sidecaption[t]
\includegraphics[width=2.8in]{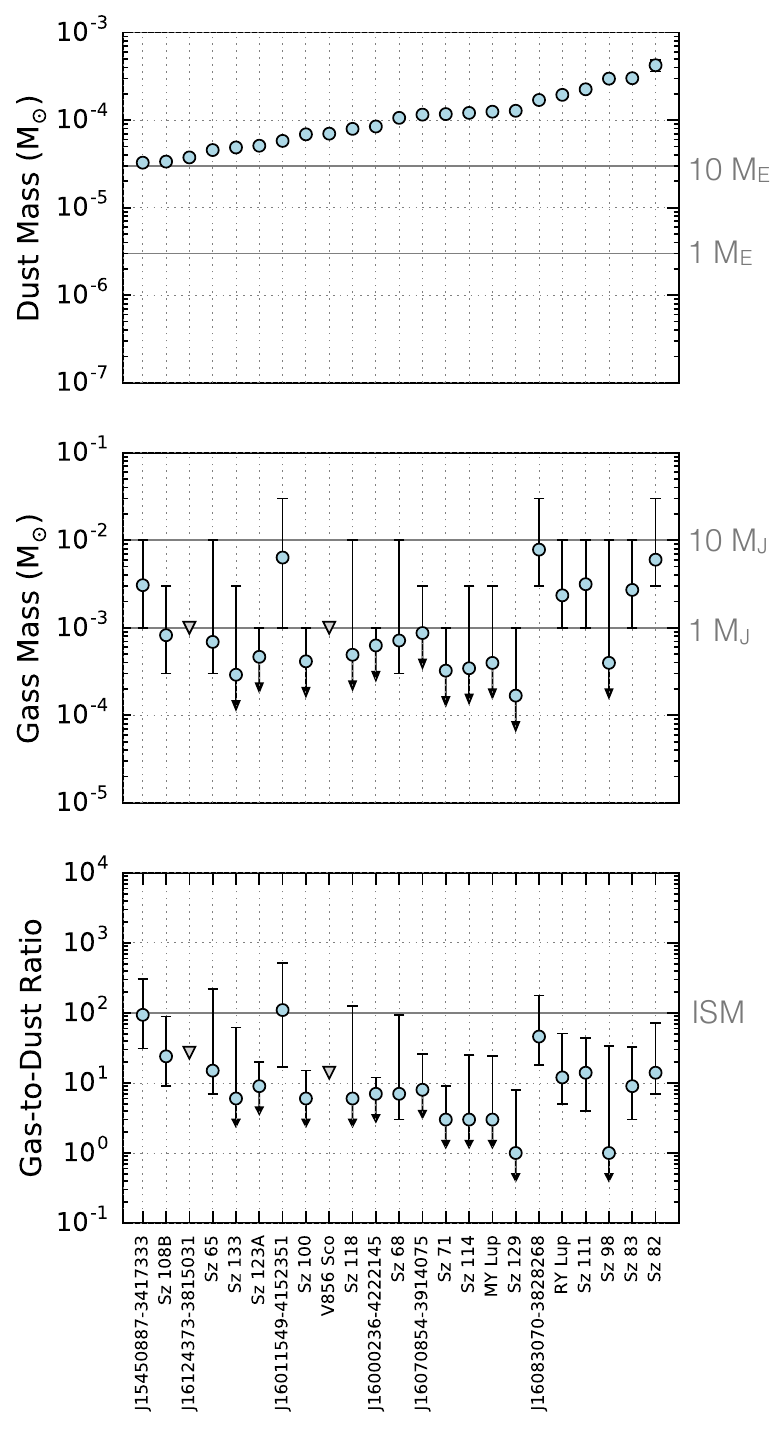}
\caption{
Disk masses derived from the Lupus ALMA survey.
The top panel shows the dust masses as derived from the
$890\,\mu$m continuum using equation X.
The middle panel shows the gas masses as derived from the
CO isotopologue emission as described in \S\ref{sec:gas_CO}
and plotted in Figure\,\ref{fig:lupus_CO_luminosities}.
The lower panel shows the gas-to-dust ratio, and how it appears
to be significantly lower than the ISM value of 100 for almost
all disks here.
This Figure is adapted from \citet{Ansdell16}
to only show 23 most massive disks from the full sample of
90 as these were all detected in \13CO\ and most in \C18O,
thereby permitting the best constraints on the gas mass.
}
\label{fig:lupus_gd_range}
\end{figure}

Some data points lie below the model locus shown here
indicating lower \C18O\ line luminosities than predicted.
This is probably due to selective photo-dissociation of this
rare isotopologue, which cannot self-shield as effectively as CO
\citep{vdb88}.
More detailed modeling that takes this into account
confirms the basic findings here \citep{Miotello16}.

The range of Lupus disk masses is plotted in Figure~\ref{fig:lupus_gd_range}
where the dust mass is derived following the procedure described in
\S\ref{sec:dust} and the gas masses from the CO isotopologues.
The most massive dusty disks, $M_{\rm dust}>10\,M_\oplus$, are shown here
\citep[the full sample is shown in][]{Ansdell16}.
All are detected in \13CO\ and many in \C18O, but in general the lines
are quite weak and the implied gas masses are low,
typically $M_{\rm gas}\sim 1\,M_{\rm Jup}$.
This mirrors the \citet{Williams14a} results in Taurus which itself has
precedent in the low \13CO\ line flux found by \citet{Kastner97} in TW Hya.
The inferred gas-to-dust mass ratios are plotted in the lower panel.
The high dust and low gas masses implies ratios that are generally
much lower than the fiducial ISM value of 100.

Just as molecular cloud and core masses derived from CO observations
depend inversely on the molecular abundance, so the disk gas masses
shown in this section are scaled to an assumed value,
[\twelveCO]/[H$_2$]$=10^{-4}$.
This value has been calibrated in clouds and cores but there
are few direct measurements in disks and they disagree.
The discrepancy is most significant with the HD measurements
described in \S\ref{sec:gas_HD}.
Ultimately, the CO abundance is the largest source of uncertainty in
gas mass measurements from CO and, because of the relative
ease of measurement of these lines,
this is  a critical issue for disk studies.
We discuss this further in \S\ref{sec:reconciliation}.

\subsubsection{Gas Surface Density Profiles}
The modeling methodology used in Figure~\ref{fig:lupus_CO_luminosities}
to measure bulk gas masses can be extended to determine
gas surface density profiles.
The basic idea, described in detail in \citet{Williams16},
is to compare a library of model images with a resolved line map.

\begin{figure}
\sidecaption
\includegraphics[width=4.7in]{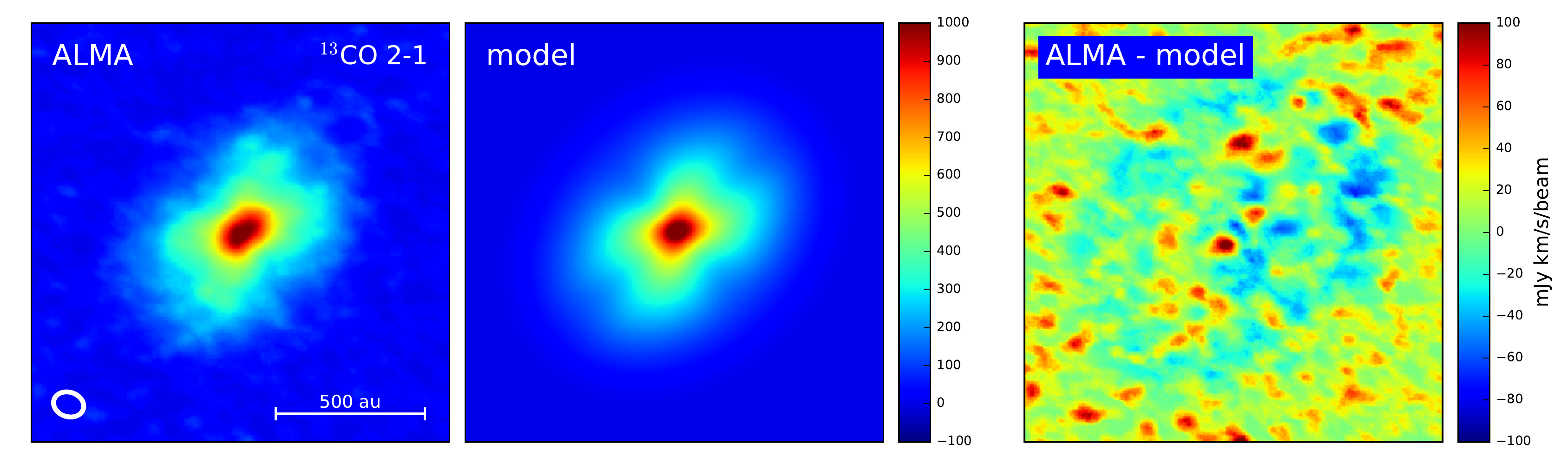}
\caption{Comparison of \13CO\ 2--1 integrated intensity images for the HD\,163296 disk.
The left panel shows the ALMA map,
the central panel the model fit for the best fit
model parameters, and the right panel shows the difference
between the two.
The colorbars show the range of intensities for each map;
the ALMA and model images on the same scale from
$-100$ to 1000\,mJy\,\kms\,beam\e,
and the difference image from $-100$ to 100\,mJy\,\kms\,beam\e.}
\label{fig:hd163296}
\end{figure}

Figure~\ref{fig:hd163296} shows that this works very well
in the case of the large, bright disk around the Herbig Ae
star, HD\,163296.
In this case, the \13CO\ map is well resolved, has a high
signal-to-noise ratio, and the
temperature structure of the disk was well constrained
by modeling the CO data independently \citep{Rosenfeld13}.

It will be interesting to carry out similar modeling of
other resolved \13CO\ line images with different gas-to-dust
(or CO line to continuum) ratios. This requires a
similar dynamic range in both spatial and intensity scales.
\citet{Williams16} show that moderate resolution ($0\farcs2-0\farcs3$)
ALMA observations of a few to tens of minutes (mass dependent)
suffice to measure the gas profiles of typical disks
found around T Tauri stars.

The $J=2-1$ transition is ideally suited for this analysis
as the CO, \13CO, and \C18O\ lines can all be observed
simultaneously with ALMA and at the same resolution and
sensitivity. A combined modeling of the CO and \13CO\
data allows the temperature and density structure to be
derived in tandem. By then comparing to the \C18O\ map,
the isotopologue ratio can be determined and any
radial variation thereof. This provides an interesting
test of selective photo-dissociation models which have
been invoked for explaining the oxygen isotope variation
in the Solar System \citep{McKeegan11}.
As the 1.3\,mm dust continuum image is also obtained,
this a powerful combination for studying the distribution
of the solid and gas components.

This can readily be extended to other molecules.
Even if absolute abundances are uncertain, resolved maps can
show the radial dependence of relative abundances very well.
As the dust processing and chemical timescales vary with radius,
we can test models that predict different abundance patterns.

\subsection{Comparison of CO and HD}
In some instances there have been attempts to measure the CO abundance in disks which provide information on its viability as a mass tracer.  This is complicated by the fact that CO is known to be frozen as ice in the dense midplane as discussed in \S~3.4 \citep{dgg97, aikawa_vanz02, bergin_ppv,semenov12}.  This facet is spectacularly confirmed by recent imaging of the HD~163296 disk \citep{Rosenfeld13}.
In the right-hand panel of Fig.~\ref{fig:abuncal} we provide the limited set of CO abundance measurements in disk systems, which illustrate the complexity of this problem.  Thus \citet{Schwarz16} combine the detection of HD with mulitransitional $^{13}$CO and C$^{18}$O ALMA observations of TW Hya to resolve the CO abundance structure in TW Hya \citep[see also][]{Nomura16}; a disk average value presented in \citet{favre13a}.   The assumption in both of these cases is that the HD emission is strongly curtailed for T $<$ 20 K  where CO is frozen out as ice.  Hence the CO/HD column density ratio is not affected by freeze-out and traces the abundance in the emissive layers.  This is borne out by excitation calculations \citep[e.g.][]{favre13a}.   With ALMA the disk thermal structure is resolved and the effect of temperature on the HD emission can be mitigated.   In all cases, both within the CO snowline (near 20 AU) and beyond the CO snowline the derived CO abundance is well below the ISM value.  This is argued to be due to incorporation of CO into solids either as CO$_2$ or perhaps processed into more chemically complex forms 
\citep{bergin14, Furuya14, Reboussin15, Yu16, Eistrup16}.

In two cases both CO and H$_2$ have been observed in absorption in the UV towards RW Aur \citep{France14}  and AA Tau \citep{France12a}.  In one case the derived abundance is very close to interstellar (RW Aur), but in the case of AA Tau the CO abundance measurement is 0.4 relative to H (i.e. four orders of magnitude {\em above} interstellar).    UV observations generally probe gas very close to the star and/or at high altitudes in the disk (with low column and UV exposure, e.g. $A_V \ll 1$).   Thus it is difficult to extrapolate these numbers to the overall disk mass; regardless, they do represent useful information.   These few measurements represent the only abundance estimates that are done independent of the dust.  The remaining values come from a range of studies that infer the H$_2$ mass/column from dust observations via spectral energy distribution (SED) modeling or through resolved thermal continuum emission maps.  In one study towards BP Tau, \citet{Dutrey03} observed CO isotopologue emission with the IRAM Plateu du Bure interferometer and the 30m telescope.   They argue that the  overall disk is warm and above the CO sublimation temperature.  Assuming a normal gas/dust mass ratio of 100 they derive a CO abundance (relative to H$_2$) more than two orders of magnitude below the ISM value of $\sim 9 \times 10^{-5}$ (dashed line in Fig.\ref{fig:abuncal}).   In HD 100546, \citet{Kama16b} and \citet{Bruderer12} use numerous molecular and atomic lines to explore the chemistry in this disk in the framework of a 2D thermochemical model.     In all, if the gas/dust mass ratio is 100, they find a reduced overall amount of elemental carbon and therefore a reduced CO abundance.

In \S\,3.4, we discussed the comparison of C$^{18}$O emission and dust in the
\citet{Williams14a} and \citet{Ansdell16} surveys.
Here the problem of abundance vs mass is clearly revealed.
If the gas-to-dust mass ratio is 100, there is a range of CO abundances that span
nearly 2 orders of magnitude.
If the CO abundance is $\sim 10^{-4}$ as in the ISM,
then the gas-to-dust mass ratio ranges from $\sim 1-10$.
Thus the issue is revealed -- is this evolution of the CO abundance or is this
gas dissipation (or some combination of both).
Either issue represents a  fundamental statement about the overall physical and
chemical evolution of disk systems.

\section{Reconcilation?}
\label{sec:reconciliation}

In sum we have two disparate sets of measurements.   Observations of a statistically significant sample of disks studied in the emission of CO isotpologues find that they are missing CO emission relative to measured dust masses.  Recall that the dust mass measurements are a lower limit to the total solid component as grains much larger than the observing wavelength have little emission. Thus the low CO emission is not due to an overestimate of the solids. It can be interpreted as either
missing carbon monoxide assuming that H$_2$ is still present
or as very low gas/dust mass ratios (i.e. missing gas mass).
The magnitude of these effects (at least two orders of magnitude)
is effectively summarized in the right hand panel of Fig.~\ref{fig:abuncal}.
We discuss the chemical and then the physical possibilities in turn.

There are three disks with gas mass  measurements from HD \citep{bergin_hd, McClure16} and, in the case of TW Hya and GM Aur (to a lesser extent DM Tau), this suggests that CO is missing while hydrogen gas is abundant \citep{favre13a, Schwarz16, Nomura16, Kama16a}.   
It is important to stress that these inferences are not limited to CO and its isotopologues alone.     Surveys of [\ion{O}{i}] 63$\mu$m emission in Taurus find that, for normal gas/dust mass ratios ($\sim$100), models tend to over-predict line emission \citep{Aresu14}.   Similarly, the ground state lines of water vapor at 557 and 1113 GHz have been detected in only one T Tauri disk system \citep[TW Hya;][]{hoger11a}, with non detections in over 13 potentially gas-rich disk systems \citep{Du16}.  Detailed models  predict that these lines would be too strong by large factors and require a reduced water abundance by two orders of magnitude under the assumption of a ``normal'' gas/dust ratio \citep{Bergin10b, hoger11a, Du16}, but see also \citet{Kamp13}.   In the case of TW Hya, \citet{Du15} performed detailed thermochemical modeling of CO isotopologue,   [\ion{O}{i}], OH, HD, and water vapor emission demonstrating that volatiles appear to be depleted in the upper atmosphere of the disk.   Thus there appears to be a systematic effect that affects the main volatile carriers of at least carbon and oxygen.\footnote{Recently, there has been the suggestion that CO is missing from the gas phase in regions in close proximity to young protostars \citep{Anderl16}.  Perhaps hinting that these effects begin during the collapse phase.}

One interesting chemical effect that would aid in reconciliation, at least for CO, is isotopic self-shielding.  It is well known that the photodissociation of carbon monoxide proceeds via a line process which means that the presence of CO molecules along the line of sight toward the radiation source can shield other molecules downstream.  This ``self-shielding'' effect has an innate column density dependence; hence $^{12}$CO is more efficiently shielded than less abundant $^{13}$CO and so on.   Since we are mainly using C$^{18}$O as our preferred H$_2$ mass tracer it could probe less mass in the disk than say HD, which would result in a reduced abundance measurement. \citet{Miotello14} and \citet{Miotello16} explore these effects in detail and show that a CO abundance reduction of nearly an order of magnitude could be accounted via this mechanism.  However, HD also self-shields and hence it traces a similarly reduced column of H$_2$ \citep[see][]{bergin14}.  Furthermore,  \citet{Schwarz16} use ALMA observations of CO isotopologues towards TW Hya to constrain the thermal structure and, with HD, the CO abundance.    They find that the abundance is reduced by nearly two orders of magnitude.   While, self-shielding works in the right direction, it cannot account for the observed magnitude of the abundance depletion in TW Hya at least, nor does it account for the presence of similar effects seen in oxygen-bearing molecules.

An additional chemical scenario is that we are witnessing the formation of icy planetesimals or at least pebbles that are trapped in the midplane \citep{Dutrey03, Chapillon10, Bergin10b, hoger11a, Du15, Kama16a}.  A wide variety of observations have now shown that the \emph{dust} mass in disks is radially and vertically stratified due to the combined effects of coagulation, settling, and drift \citep[see discussion in][]{Andrews15}.   Molecular ices will preferentially form or freeze-out onto small dust grains (as they carry the surface area) as gas dynamically mixes to cold layers.
Through sticking and grain growth these ice coated small grains will incorporate volatiles into larger grains that become trapped in the dust-rich midplane.   This so-called ``vertical cold finger effect''  \citep{Meijerink09, Kama16a} can deplete the upper emissive layers of volatiles, particularly water with its relatively high sublimation temperature \citep{Krijt16}.   Due to its low sublimation temperature ($\sim 20 - 30$~K) the sequestration of CO in ices is less certain.  However, a number of authors  have independently characterized how the carbon in CO can be reprocessed into less volatile forms such as CO$_2$ and organics
\citep{bergin14, Furuya14, Reboussin15, Yu16, Eistrup16}.   For CO the reprocessing is linked to the presence of X-rays or UV photons along with active surface chemistry and hence will be most active on surface layers and in the outer disk. 

\begin{figure}[b]
\sidecaption[t]
\includegraphics[width=2.9in]{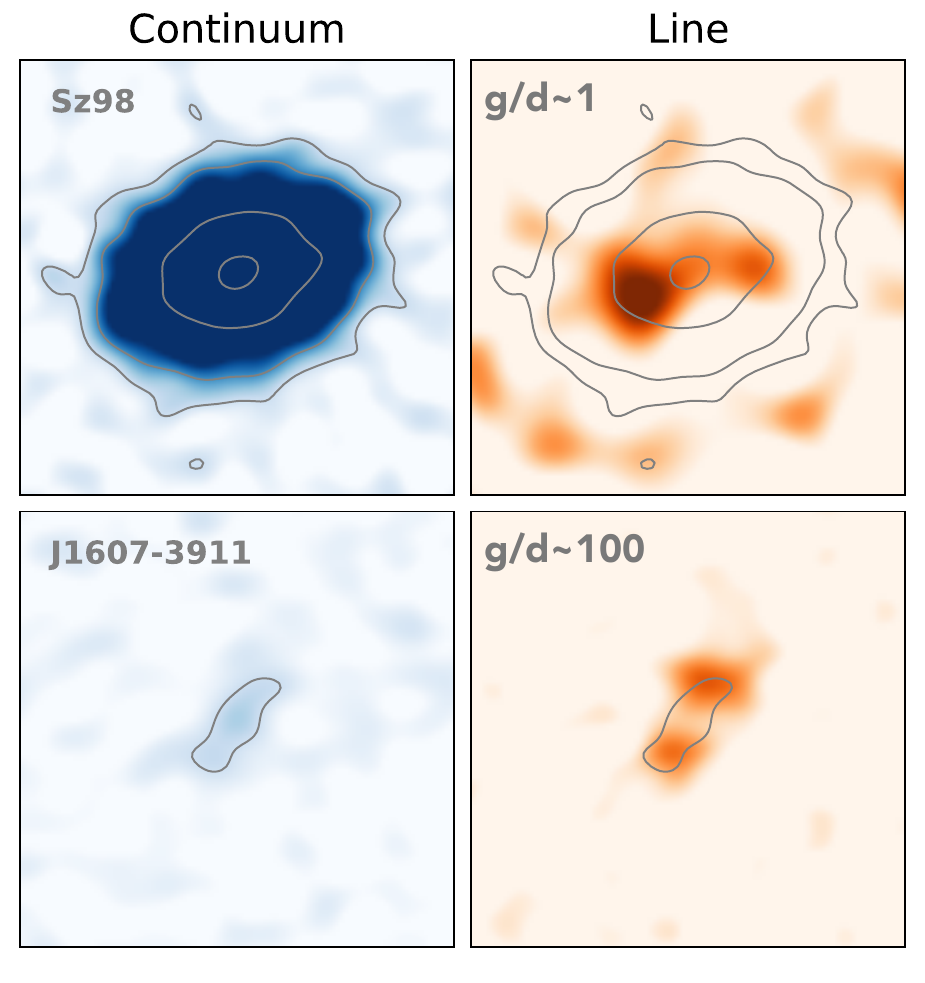}
\caption{
The two most extreme gas-to-dust ratios in the Lupus disk survey.
The left column shows the continuum emission on the same scale 
(1-10 mJy\,beam\e)
for two different disks, the large, bright Sz\,98 on the top and
the low mass J1607-3911 on the bottom. The right column shows the
integrated \13CO\ 3--2 emission for each disk on the same scale
(40-400\,mJy\,\kms\,beam\e)
and with the dust contours overlaid.
It is clear that the \13CO\ line-to-continuum ratio is very different 
in the two sources which suggests either very divergent chemistry or
strikingly different gas-to-dust ratios.
}
\label{fig:gas2dust_extremes}
\end{figure}

% Planetary synthesis models inevitably form gas giants in long-lived
% gas-rich disks \citep{mordasini12}.
% The exoplanetary record tells us then that the gas
% eventually dissipates: there are too many Neptunes and not enough Jupiters!
% We need more measurements of disk gas content and evolution.
% The HD emission is generally very weak, both because HD is very rare and most of
% the disk is too cold to populate its excited state.
% Furthermore, the far-infrared line could only be observed with the space-based
% Herschel observatory, which is no longer operational. Consequently, there are only
% three HD detections in disks and a handful of other upper limits.
% Millimeter wavelength CO rotational lines can be measured from the ground, however,
% so many disks can be surveyed and imaged at sub-arcsecond resolution.

Alternately, the results to date show that most, but not all, disks have weak CO isotopologue emission, while HD (due to the limited lifetime of Herschel) has only been detected in 3 systems.   Crucially the HD line is yet to be spectrally resolved and thus might contain contributions from unrecognized components such as outfows (jets or winds). 
Furthermore, the statistics show that, whatever the cause, the CO line-to-continuum ratio
varies widely from disk to disk within a single star-forming region
(Figure~\ref{fig:gas2dust_extremes}).
This is a challenge for chemical sequestration models as it is unclear why they
should be 99\% efficient in some cases but 0\% in others with the same age.
In a sense this mirrors the same issues for disks with respect to the fact that,
for a given age, disks surrounding similar mass stars exhibit a range of infrared excesses and dust masses (\S~2.1),
which must relate to dust evolution.

If chemistry is not the predominant cause and  the [CO]/[H$_2$] abundance in
the warm molecular layer is indeed similar to the ISM,
then the conclusion would be that the gas is 
indeed preferentially lost relative to dust.
In the HD 163296 example shown in Figure~\ref{fig:hd163296_gas_dust} the gas at large radii and scale heights is dust poor due to the effects of settling and drift.
These are the same regions in a disk that are less strongly bound to the star,
and which can be lost through photo-evaporation, possibly assisted by
magnetic fields \citep{AlexanderR14, Gressel15}.

Photoevaporative flows from disk surfaces depend strongly
on the radiation field, whether EUV, FUV, or X-ray \citep{AlexanderR14}.
These purely hydrodynamic flows are generally strongest at small radii
but the effect on the gas-to-dust has not yet been modeled.
There is a direct measurement of a photoevaporative flow in one case
(TW Hydra again) through mid-infrared spectroscopy of the [NeII],
but the mass loss rate, $\sim 10^{-10}\,M_\odot\,{\rm yr}^{-1}$,
is too low to significantly affect the disk composition on Myr timescales
\citep{Pascucci11}.

There is much new work on angular momentum transport in disks,
recently summarized by \citet{Turner14}.
Just as advances in instrumentation allow new observational
discoveries, so faster computers allows more detailed modeling
and better treatment of non-ideal magneto-hydrodynamical effects.
The ionization fraction, and therefore the coupling between the gas and
dust with the magnetic field, vary widely within a disk.
The dense disk mid-plane may be so shielded from radiation
and cosmic rays \citep{Cleeves13a} that the coupling may be too weak for
the magneto-rotational instability \citep[MRI;][]{balbus91}
to drive accretion onto the star.
The surface layers are more highly ionized and tied to the magnetic field.
The MRI may still operate in these layers and drive laminar accretion flows
but disk winds that are launched at large radii may be more effective
in re-distributing the angular momentum.

Such global disk winds are predicted to have mass loss rates
$\sim 10^{-8}\,M_\odot\,{\rm yr}^{-1}$
and can therefore remove $10\,M_{\rm Jup}$ in $\sim 1$\,Myr
from the disk surfaces at large radii \citep{Bai16}.
This is greater than the median disk mass around a solar mass star in
Taurus for an initial gas-to-dust ratio of 100.
Although there are as yet no explicit calculations of the effect on
the gas-to-dust ratio, it seems that disk winds can potentially rapidly
remove disk gas atmospheres but leave millimeter grains in the midplane
largely unaffected. As with accretion, the winds could be variable
and vary greatly from disk to disk due to the intricacies of the initial 
magnetic field geometry including its polarity with respect to the
rotation axis.

ALMA observations show kinematically coherent CO structures
around disks on $\sim 2000$\,au scales that may be
large scale winds \citep{Klaassen13, Klaassen16}.
The flows have low speeds similar to Keplerian rotation rates at tens of au,
and do not appear to be swept up material at the edges of powerful
jets created at the star-disk interface.
Mass loss rates are estimated to be
a few $10^{-8}\,M_\odot\,{\rm yr}^{-1}$.
Similarly large scale features have also been seen around FU\,Orionis objects.
These are very young, low mass protostars that have undergone a recent
(within a few decades) extreme luminosity outburst that is
thought to be due to a large, stochastic accretion event.
\citet{Zurlo16} and Ruiz-Rodriguez (submitted)
find rings and bow-shock shaped features in CO images that
appear to be the lit-up edges of very wide outflow cavities.
In one case, HBC\,484, the cavity opening angle is so wide,
$\sim 150^\circ$, that it would intersect the disk surface.
The observations are more suggestive than definitive at this
stage but, if most stars undergo similarly eruptive events
early in their history, such outbursts could potentially blow away
disk atmospheres and drastically alter their composition.
The outbursts and their effects are stochastic so we
might expect a wide range in gas-to-dust ratios by the 
time that the stars become optically visible Class II protostars.
The lost disk gas would be distributed on large angular scales
and ultimately photo-dissociated.  

Finally, as a sign of reconciliation between the two authors,
we note that we we may be witnessing both of these effects,
chemical sequestration and gas loss, simultaneously.

\section{Conclusion and future prospects}
\label{sec:conclusion}

Reconciliation between these disparate mass measurements remains possible.
In the near term, within the next 2 years, the HIRMES (High Resolution Mid-infrared Spectrometer) will fly on the Stratospheric Observatory For Infrared Astronomy.  HIRMES will be able to detect HD towards TW Hya and spectrally resolve the line.   While TW Hya is nearly edge on with an inclination $\sim$5-7$^\circ$, with little velocity structure beyond 20 AU, HIRMES will certainly be capable of determining if the HD line originates in whole or partially from a jet, wind, or from the disk.    In the more distant future (late 2020's, 2030's) concepts such as SPICA (Space Infrared Telescope for Cosmology and Astrophysics; JAXA) or the Origins Space Telescope (NASA) could survey HD emission towards hundreds of disk systems and detect lines at very low levels of flux density ($\sim 10^{-21}$ to $10^{-20}$ W/m$^2$).    This would provide a statistically significant sample that can be compared to the extensive archive of ALMA data that will exist by that time.  In this regard, JWST has the ability to detect the pure rotational lines of H$_2$, that is provided there is a significantly warmer H$_2$ layer above the optically thick dust concentrated towards the midplane.   This might prove useful in constraining the H$_2$ surface density in the very innermost disk radii.

Direct mass measurements, or at least useful constraints, may also be possible through
ultra-high resolution ALMA imaging of dust structure.
If the ratio of disk to stellar mass exceeds $\sim 20$\%, the gravity of the disk is
strong enough to produce detectable features \citep{Cossins10, Dipierro14}.
Azimuthally symmetric rings are
found in HL Tau and TW Hya \citep[][respectively]{HLTau_ALMA, Andrews16},
but spirals that may indicate self-gravity have also been found \citep{Perez16}.
A large imaging survey is underway that will show the range and occurrence of
various morphologies, and that can be compared to mass determinations from spectral lines.

More generally, both suggested scenarios, gas dissipation and chemical sequestration, likely have attendant chemical effects that might be revealed via ALMA molecular line emission studies.    As an example, it has been suggested that strong hydrocarbon emission seen in several disks might be related to dust evolution and the chemical sequestration of volatiles \citep{Kama16b, Bergin16}.  At face value, if true, then these chemical effects (in this case hydrocarbon production) should track with the overall amount of CO present which may have strong radial dependencies.  Along these lines, additional chemical signatures might be found in models of the gas  chemistry in dissipating gaseous disk systems.   Another method would be the inference of the gas density via excitation analyses using a molecule that traces close to the dense midplane in a specific location.  One example could be  N$_2$H$^+$ \citep{Qi13a} which is known to show a ring structure.  The inference of an H$_2$ gas density in a specific location can be used to pin down and constrain the overall gas density structure - such modeling would at least set some limits on the overall gas mass.

The determination of disk gas mass has been a long-term problem for decades  and solving this issue will require efforts from multiple directions.  However, the effects we are discussing in this manuscript encompass nearly two orders of magnitude either in gas mass or molecular abundance; or less if it is both.   Such a large effect will have attendant signatures and in the coming years we are confident that this issue will be resolved, thereby  making a fundamental contribution to our understanding of the physics and chemistry of planet formation.

%%% REFERENCES
% (This doesn't work...)
%\bibliographystyle{apj}
%\bibliography{refs,apj-jour}

% kludge for now...
\newcommand{\nat}{{ Nature }}
\newcommand{\aap}{{Astron. \& Astrophys. }}
\newcommand{\aj}{{ Astron.~J. }}
\newcommand{\apj}{{ Astrophys.~J. }}
\newcommand{\araa}{{Ann. Rev. Astron. Astrophys. }}
\newcommand{\apjl}{{Astrophys.~J.~Letters }}
\newcommand{\apjs}{{Astrophys.~J.~Suppl. }}
\newcommand{\apss}{{Astrophys.~Space~Sci. }}
\newcommand{\icarus}{{Icarus }}
\newcommand{\mnras}{{MNRAS }}
\newcommand{\pasp}{{ Pub. Astron. Soc. Pacific }}
\newcommand{\ssr}{{Space Sci. Rev.}}
\newcommand{\planss}{{Plan. Space Sci. }}
\newcommand{\physrep}{{ Phys. Rep.}}
\newcommand{\bain}{{Bull.~Astron.~Inst.~Netherlands }}
%\begin{thebibliography}{}
%\input{refs}
%\end{thebibliography}
\bibliographystyle{aps-nameyear}
\bibliography{ted+jpw}

\end{document}